\RequirePackage{fix-cm}

\documentclass{svjour3}

\smartqed

\usepackage{amsmath}
\usepackage{framed}
\usepackage{graphicx}
\usepackage{subfig}

\usepackage{natbib}
\usepackage{hyperref}
\hypersetup{colorlinks,citecolor=blue,linkcolor=blue,urlcolor=black}

\journalname{Journal of Ambient Intelligence and Humanized Computing}

\newcommand*{\nolink}[1]{%
	\begin{NoHyper}#1\end{NoHyper}%
}

\newcommand*{\mycitep}[1]{%
	(\nolink{\citeauthor{#1}}~\citeyear{#1})%
}

\newcommand*{\mycitepp}[2]{%
	(\nolink{\citeauthor{#1}}~\citeyear{#1};~\nolink{\citeauthor{#2}}~\citeyear{#2})%
}

\begin{document}

\title{Dynamic management of a deep learning-based anomaly detection system for 5G networks}
\titlerunning{Dynamic management of a deep learning-based anomaly detection system in 5G}

\author{Lorenzo Fern\'andez Maim\'o \and Alberto Huertas Celdr\'an \and Manuel Gil P\'erez \and F\'elix~J. Garc\'ia Clemente \and Gregorio Mart\'inez P\'erez}
\authorrunning{L. Fern\'andez Maim\'o et al}

\institute{Lorenzo Fern\'andez Maim\'o \at Departamento de Ingenier\'ia y Tecnolog\'ia de Computadores, University of Murcia, 30071 Murcia, Spain\\\email{lfmaimo@um.es}\\ORCID:0000-0003-2027-4239
	\and Alberto Huertas Celdr\'an \at Departamento de Ingenier\'ia de la Informaci\'on y las Comunicaciones, University of Murcia, 30071 Murcia, Spain\\\email{alberto.huertas@um.es}
	\and Manuel Gil P\'erez \at Departamento de Ingenier\'ia de la Informaci\'on y las Comunicaciones, University of Murcia, 30071 Murcia, Spain\\\email{mgilperez@um.es}
	\and F\'elix J. Garc\'ia Clemente \at Departamento de Ingenier\'ia y Tecnolog\'ia de Computadores, University of Murcia, 30071 Murcia, Spain\\\email{fgarcia@um.es}
	\and Gregorio Mart\'inez P\'erez \at Departamento de Ingenier\'ia de la Informaci\'on y las Comunicaciones, University of Murcia, 30071 Murcia, Spain\\\email{gregorio@um.es}
}

\date{Received: date / Accepted: date}

\maketitle

\begin{abstract}
	Fog and mobile edge computing (MEC) will play a key role in the upcoming fifth generation (5G) mobile networks to support decentralized applications, data analytics and management into the network itself by using a highly distributed compute model. Furthermore, increasing attention is paid to providing user-centric cybersecurity solutions, which particularly require collecting, processing and analyzing significantly large amount of data traffic and huge number of network connections in 5G networks. In this regard, this paper proposes a MEC-oriented solution in 5G mobile networks to detect network anomalies in real-time and in autonomic way. Our proposal uses deep learning techniques to analyze network flows and to detect network anomalies. Moreover, it uses policies in order to provide an efficient and dynamic management system of the computing resources used in the anomaly detection process. The paper presents relevant aspects of the deployment of the proposal and experimental results to show its performance.

\end{abstract}


\section{Introduction}
\label{sec:intro}

Fog and Mobile Edge Computing (MEC) is a new network architecture paradigm based on a disseminated computing infrastructure that provides a service environment and cloud computing capabilities at the edge of the mobile network. MEC can be seen as a natural development in the evolution of mobile networks, where applications and services are handled within the Radio Access Network (RAN) and in close proximity to User Equipments (UE). The aim of MEC is to enable low-latency services and delay-sensitive applications ensuring highly-efficient network operation and service delivery, as well as offering an improved user experience.

MEC is recognized as an essential element of emerging technologies for the upcoming fifth generation (5G) mobile technology together with Network Functions Virtualization (NFV) and Software Defined Networking (SDN). The new 5G scenarios are characterized by several Key Performance Indicators (KPIs), as defined by the 5G Infraestructure Public Private Partnership (5G-PPP \citeyear{5GPPP:kpi}). Among these indicators, the number of connected devices (from 10 to 100 times more than 4G/LTE), the volume of mobile data per geographical area (1000 times higher), the end-to-end latency (less than 1ms), and ubiquitous 5G access including low density areas are some of the most relevant aspects that influence the evolution of current networks towards future mobile networks strengthened by the 5G technology. MEC contributes to allowing 5G to show its real potential and satisfying these demanding requirements in terms of expected performance, scalability, latency, and automation.


Furthermore, increasing attention is paid to providing user-centric cybersecurity solutions, which particularly require scalable solutions that allow collecting, processing and analyzing significantly large amount of data traffic and huge number of network connections in 5G networks. Existing solutions based on Intrusion Detection Systems (IDS) include (pro-)active approaches to anticipate and remove vulnerabilities in computing systems with which to trigger reactive actions for mitigation. However, and due to the increase of bandwidth, IDS-based solutions that make use of deep analysis techniques have been forced to evolve towards new ways of detection. These solutions have moved from inspecting raw network packets to analyzing traffic network flows with innovative AI-based techniques \mycitep{buczak2016ml}. Nonetheless, these solutions appear to be insufficient for the 5G networks, where existing detection procedures have become obsolete and, therefore, new methodologies and technologies are required. These technologies should offer, among others, the following features in an efficient, automatic and seamless way: resource management according to the number of users and their generated traffic, hot upgrade of detection models, dynamic deployment of new resources on demand, and deployment of specific analysis tools to extract detailed information.

To face this challenge, this paper substantially extends a preliminary solution in which an initial 5G-oriented architecture was presented \mycitep{loren2018access}, in order to identify cyber threats in 5G networks by making use of deep learning techniques. In this extension, our main contribution consists in extending our architecture with new modules and functionality in order to provide a MEC-oriented solution to detect network anomalies in real-time and autonomic way. Our proposal uses deep learning techniques to analyze network flows and to detect network anomalies. Furthermore, it uses policies in order to provide an efficient and dynamic management system of the computing resources used in the detection process. Several families of management policies have been defined with the goal of making decisions to deploy and control the network infrastructure, the deep learning framework, and the network services running on the edge of the network. In addition, this contribution presents relevant aspects of the deployment of the proposal and experimental results to show its performance. These experiments are focused not only on the detection precision of the deep network but also on the CPU time consumed by the proposed deep architecture during classification. An exhaustive set of execution time measurements were carried out in order to model the throughput of our anomaly detector. This runtime information is used in policies that define actions to dynamically self-adapt our system to the traffic load requirements.

The remainder of the current paper is structured as follows. Section~\ref{sec:related_work} discusses related work on anomaly detection systems, MEC-oriented security solutions and deep learning proposals. Section~\ref{sec:architecture} describes the MEC-oriented architecture. Section~\ref{sec:usecase} presents a realistic use case where several concerns of 5G networks are depicted. Management policies are described in Section~\ref{sec:policies} and the orchestration process for the use case is outlined in Section~\ref{sec:orchestration}. Section~\ref{sec:performance} details the deployment using deep-learning techniques and the performance results. Finally, conclusions are drawn and future work is outlined in Section~\ref{sec:conclusions}.


\section{Related work}
\label{sec:related_work}

In the past few years, several solutions have been proposed to detect and mitigate attacks in computer networks by considering the Software Defined Networking (SDN) paradigm and Network Function Virtualization (NFV) techniques. In this context, a list of potential cyber threats in current networks was proposed in \mycitep{mantas:5g:2015}, which could also be applied to future 5G networks. In this security context, botnets are highlighted as one of the most powerful cyber threats for 5G networks \mycitep{anagnostopoulos:ijis:2016}. BotGuard was presented in \mycitep{chen:wujns:2017} as a framework for real-time botnet detection in SDN networks. The proposed architecture extracts the botnets' most significant features to be able to detect Command and Control (C\&C) channels. Another solution was proposed to detect and mitigate botnet attacks in 5G networks (Gil P\'erez et al \citeyear{gil:ic:2017}). This proposal involves the tight combination of both NFV and SDN technologies to provide effective detection and mitigation of cyber-attacks in 5G networks. Finally, an architecture that combines NFV and SDN features was proposed in \mycitep{cleder:iscc:2016} to create sophisticated network resilience strategies. This architecture uses a control-loop to monitor and analyze the network state, indicating if specific parts can be reconfigured or replaced to improve the detection capabilities.

Regarding anomaly detection, many approaches have been explored \mycitepp{chandola2009}{garcia2009anomaly} whereas the machine learning perspective has received special attention in recent years. For example, an FPGA architecture based on a Block-Based Neural Network (BBNN) for an anomaly-based IDS is proposed in \mycitep{tran2012bbnn}. In particular, a complete survey dealing with solutions to quickly classify collected network flows and detect attacks, or malicious code, is provided in \mycitep{gardiner2016sml}. Other recent works based on deep learning have provided state-of-the-art classification performance \mycitepp{yin2017}{chang2017}. Additionally, deep learning has the ability of automatically extracting high-level features from large amounts of raw data, preventing overfitting thanks to recent regularization techniques (Wang et al \citeyear{wang2017hast}). This is in contrast to classic machine learning classification algorithms that typically rely upon feature engineering methods to reduce the dimensionality of the input \mycitep{chang2016rank}.

Despite the progress made by the previous proposals, none of them considers the anomaly detection in 5G networks, nor the efficient and dynamic management of the network infrastructure and resources used in the detection processes.


\section{MEC-oriented architecture for network anomaly detection}
\label{sec:architecture}

We propose an architecture oriented to MEC for detection of network anomalies in 5G networks. This architecture is based on decentralized applications, data analytics and management into the network itself using a highly distributed compute model.

\begin{figure}[hbtp]
	\centerline{\includegraphics[width=\columnwidth]{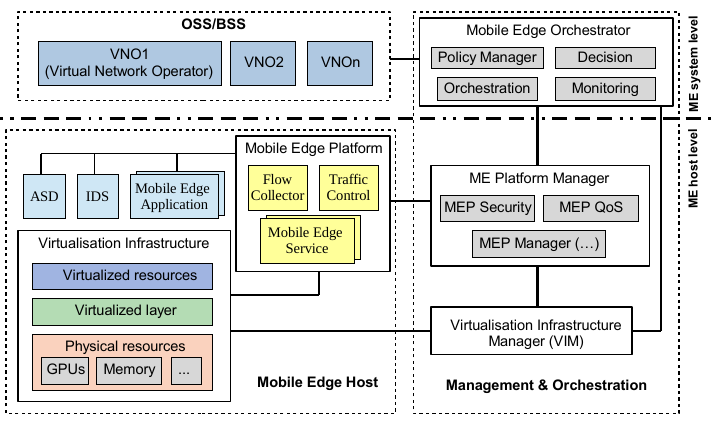}}
	\caption{High-level management and orchestration plane.}
	\label{fig:architecture}
\end{figure}

5G networks are conceived as extremely flexible infrastructures based on an architecture organized by different functional planes. These planes provide separation of duties to simplify and address a complete set of novel system requirements. Among these planes, we propose a high-level design of the management and orchestration plane, following the ETSI NFV architecture \mycitep{etsi2017nfv5G}, which has also been used as the basis for other proposals (Siddiqui et al \citeyear{siddiqui2016hierarchical}; Neves et al \citeyear{neves2017future}). MEC, like NFV, makes use of a stack of standard components, including a well-defined compute platform and virtualization layer. Therefore, the infrastructures that support MEC and NFV are similar to a great extent, and thus the same platform can be shared between both Virtual Network Functions (VNF) and MEC applications. In addition, combining the SDN paradigm with NFV techniques allows decoupling the software implementation of VNFs from the underlying hardware, providing and enhancing the flexibility in the management of the network resources \mycitep{mijumbi:NFV:2015}.

The proposed cyber defense architecture, as shown in \figurename~\ref{fig:architecture}, consists of two levels: \textit{Mobile Edge System level} and \textit{Mobile Edge Host level}. The latter includes the \textit{Mobile Edge Host}, an entity that contains a Virtualisation Infrastructure (VI) which provides storage, memory, computation, and network resources, for the purpose of running \textit{Mobile Edge Applications}. In addition, ME Host provides \textit{Mobile Edge Platform}, that is, the essential functionality required to run ME Apps on a particular VI and enable these ME Apps to provide and consume \textit{Mobile Edge Services}. The top level of our architecture is in charge of defining and controlling the general behavior of the Anomaly Detection (AD) system. To this end, this level is made up of the \textit{Operation Support System} (OSS) and the \textit{Mobile Edge Orchestrator}. The OSS deals with the logic of the AD system, whereas the ME Orchestrator controls the general behavior of the infrastructure by considering the policy set defined by Virtual Network Operators (VNO) of the OSS/BSS.

ME applications and services like IDS, flow collectors or Anomaly Symptom Detection (ASD) --described below-- are instantiated on the VI of the ME host based on configurations or requests validated by the management and orchestration plane. This plane comprises the two levels of our architecture and it is made up of three functional elements: ME Orchestrator, ME Platform Manager, and VI Manager. On the one hand, ME Orchestrator has an overview of the complete mobile edge system. On the other hand, ME Platform Manager and VI Manager handle the management of the ME-specific functionality of a particular ME host and the applications running on it.

\begin{figure}[hbtp]
	\centerline{\includegraphics[width=\columnwidth]{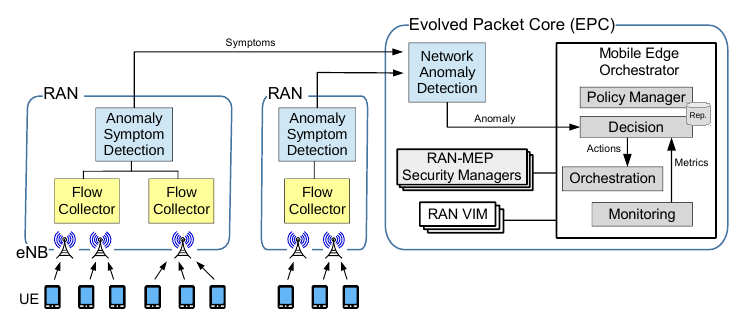}}
	\caption{Network Anomaly Detection System.}
	\label{fig:proposal}
\end{figure}

To achieve effective network anomaly detection in 5G, we propose a distributed system, depicted in \figurename~\ref{fig:proposal}, which is made up of virtualized components with three functions: Flow Collection (FC), Anomaly Symptom Detection (ASD) and Network Anomaly Detection (NAD). FC is in charge of retrieving and storing the network flows as well as providing network flow features in a way suitable to the ASD module. ASD focuses on the quick detection of anomaly symptoms examining the computed network flow features. Here, a symptom is any trace or sign of anomaly in the network traffic generated by UEs connected to the RAN. Finally, NAD is a collector of timestampted and RAN-associated symptoms, where a central process analyzes the timeline and the relationship among these symptoms to identify any network anomaly.

Once an anomaly is detected, NAD immediately informs about the type of anomaly to the ME Orchestrator. In particular, the \textit{Decision} module collects all anomaly information and combines it with metrics provided by the \textit{Monitoring} module. Among the potential metrics, the most relevant are related to the status of the resource usage (percentages of CPU and memory) and the network flows. Both VI and ME Platform managers collect monitoring data produced by their corresponding RANs and, then, they provide the Monitoring module with formatted monitoring data. The \textit{Policy Manager} module is in charge of storing policies and keeping their consistency. It also provides the Decision module with policies and contextual information to decide the best actions to perform after evaluating both the anomalies detected and the input provided by the Monitoring module.

Below we highlight some of the main actions that the \textit{Decision} module can perform to optimize both the anomaly detection processes and the usage of network resources.

\begin{itemize}
	\item \textit{Adapting virtualized resources}. When a given resource is overloaded, our solution is able to deploy new virtualized resources, change their configurations, or balance the load of flow collectors.
	\item \textit{Optimizing ASD and NAD functions}. To optimize detection processes during network traffic fluctuation, our solution can replace the deep learning framework or the detection model.
	\item \textit{Extending ASD and NAD functions}. Precise detection components may avoid false-positives. In this context, the proposed solution is able to instantiate DPI mechanisms that permit a deep search into the L2/3 flows.
\end{itemize}

At this point, it is important to note that the previous aspects allow our proposal to be highly flexible and extensible when it performs anomaly detection processes.


\section{A use case of 5G networks}
\label{sec:usecase}

This section describes a MEC-oriented use case that points out the need of managing anomaly detection and reaction capabilities of 5G networks efficiently, autonomically and in real time. To enable this management, it is critical to consider some important aspects of 5G networks like the communications latency, the number of simultaneous users consuming services, and the locations and mobility of users.

\begin{figure}[hbtp]
	\centerline{\includegraphics[width=1\columnwidth]{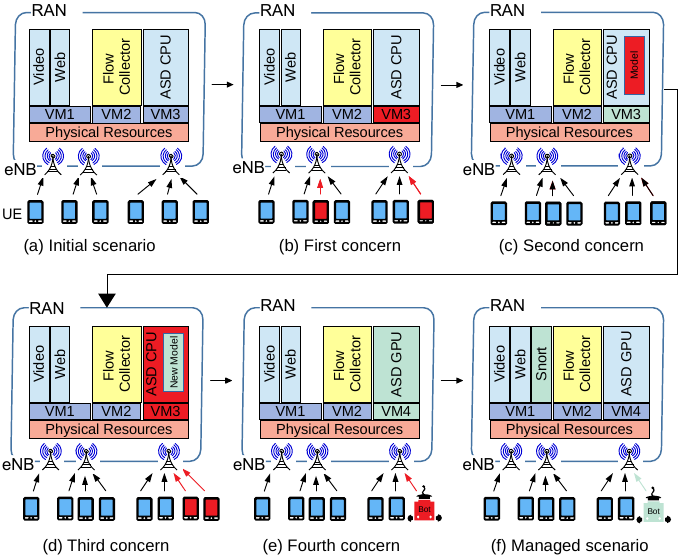}}
	\caption{Use case showing different concerns and how the proposed solution is able to manage them. Concerns are highlighted in red and their mitigation in light green.}
	\label{fig:usecase}
\end{figure}

\figurename~\ref{fig:usecase}(a) shows a scenario where mobile users are consuming different network services such as browsing, video on demand, instant messaging, e-mail, and voice over IP. These services are provided by a Radio Access Network (RAN) that ensures the low latency capability of 5G networks. The RAN has generic hardware on which virtual machines (VMs) run and provide the previous services as well as network anomaly detection and mitigation capabilities. In particular, in an initial configuration, the RAN has three VMs: VM1 provides network services, VM2 collects network flows, and VM3 analyzes them and detect anomalies by using an Anomaly Symptom Detection (ASD) module. This module requires the minimum physical resources and, therefore, it runs on a CPU to search anomaly symptom in network flows. To this end, the ASD module uses deep learning techniques. When an anomaly symptom is identified, ASD module informs the Network Anomaly Detection (NAD) module. NAD combines and analyzes the set of symptoms in order to determine the effective network anomaly detection as described in Section~\ref{sec:architecture}.

Considering the previous scenario, \figurename~\ref{fig:usecase}(b) highlights in red the moment when the number of users starts increasing in the RAN due to a given event. A larger number of users implies a significant increase of the network traffic collected and analyzed by the RAN and, therefore, the overload of the ASD module due to the need for more memory from VM3. At this point, the \textbf{first concern} of this use case arises by stressing the need of an autonomic and efficient mechanism to manage the physical resources assigned to VMs. The solution proposed in this paper is able to assign, in real time and on demand, physical resources (computation, memory, an storage) to existing VMs according to the status of the network infrastructure and its services. In this sense, \figurename~\ref{fig:usecase}(c) highlights in light green how additional RAM memory has been assigned to VM3.

Once the CPU-supported ASD module is running thanks to the assigned memory, a new detection model is released. This advanced model could include new detection capabilities or enhanced precision. It should be pointed out that training deep learning methods is a costly process because they need a large amount of data and iterations to converge. Therefore the continuous training in order to have our model updated must be executed on external computers. This is a key aspect of deep learning techniques, which have to be periodically retrained in order to detect new types of network anomalies. At this point, the \textbf{second concern} is the need to update automatically the detection model used by the ASD module. In order to deal with this concern, the proposed solution is able to re-configure the ASD module in real time by updating its detection model and configuration. In this sense, \figurename~\ref{fig:usecase}(c) highlights in red the older detection module, and on the other hand \figurename~\ref{fig:usecase}(d) highlights in light green the new detection model deployed in the ASD module.

After the second concern has been dealt with, the number of users continues increasing, which implies the overload of the ASD module. At this point, \figurename~\ref{fig:usecase}(d) shows in red color the \textbf{third concern} of this use case, which stresses the need for an automatic and efficient mechanism to manage the ASD services allocated at the edge of the network, as well as the VM where the services run. This mechanism should also be able to detect the previous situation by considering aspects like the network traffic, the number of users, and their locations. A suitable mechanism could apply two different countermeasures. The first one consists in deploying a new VM with more computational resources ---e.g. a GPU (Graphics Processing Unit) and extra memory (RAM)--- and, if necessary, a new version of the ASD module that takes advantage of the additional computational power. Surely, the GPU-supported ASD module accelerates analytics processing due to its massively parallel architecture. In this alternative, it is also important to dismantle the existing VM with the CPU-supported ASD module to guarantee the efficient management of the computational resources. On the other hand, the second alternative is to assign new computational resources (mainly, GPU and memory) to the existing VM, deploy the GPU-supported version of the ASD module in the VM, and stop the previous version of the ASD module. The selection criteria of these alternatives depend on several factors such as, among others, efficiency or deployment time. In our use case, \figurename~\ref{fig:usecase}(e) depicts in light green how the proposed solution follows the former alternative due to the fact that not all Virtualisation Infrastructure Managers (VIM) are able to increment the resources of VMs without stopping them. In this sense, network anomaly detection is supposed to be an uninterrupted process during the different management tasks.

Finally, \figurename~\ref{fig:usecase}(e) shows in red how the ASD module running on RAN identifies suspicious C\&C channels established between a given C\&C server and its recruited zombies or bots. C\&C channels are established by the C\&C server with the aim of controlling the zombies and indicating when they have to launch an attack. In this context, it is required to perform a Deep Packet Inspection (DPI) process to analyze the packets in detail and discover if the suspicious channels belong to a specific botnet or not. Here, it is important to remember that considering the high amount of packets traveling in a 5G network, it is not possible to make a DPI inspection of the whole traffic. For this reason, the proposed solution analyzes network flows instead of network packets. When some suspicious C\&C channels have been effectively detected, the \textbf{fourth concern} focuses on the necessity of deploying Virtual Network Functions (VNFs) with DPI tools to analyze the suspicious network flows in detail. To address this concern, \figurename~\ref{fig:usecase}(f) shows in light green how the proposed solution deploys and configures a Snort tool in the existing VM1 of the RAN.


\section{Management policies}
\label{sec:policies}

The proposed solution ensures the efficient and automatic management of the detection and reaction capabilities of 5G networks by using policies to that end. Among the different existing policies, we make use of management-oriented policies, which control, automatically and in real time, the network infrastructure and services as well as the Anomaly Symptom Detection (ASD) modules considered by our proposal.

Policy actions establish the behavior of the network resources according to relevant aspects like the network traffic, the detection of anomalies, the amount and mobility of users, and the current state of the network resources. Considering the scope of each policy, we classify them into three different families, which are detailed in the following subsections; namely: Virtual Infrastructure, Anomaly Detection Function, and Mobile Edge Application policies.

Our management policies comprise the following elements: the \textit{type} of policy; the network \textit{resource}, whose information is currently being managed; the \textit{metric} with which the network resources are evaluated; the \textit{location} or region where the policy will be enforced; the \textit{date} or the period of time in which the policy will be applied; and the \textit{result} or set of actions to be carried out on the network once the policy is applied. The previous elements are shown in the next policy structure:

\noindent\small
	\begin{flushleft}
		\centering
		$Type~\wedge~Resource~\wedge~Metric~[\wedge~Location]~[\wedge~Date]~\rightarrow~Result$
	\end{flushleft}
\normalsize

\subsection{Virtual Infrastructure Policy}

The Virtual Infrastructure (VI) policies aim at configuring virtual network infrastructure at the edge of the network to ensure that applications are running efficiently. Among the possible actions enforced by this kind of policies, we highlight the configuration of physical/virtual resources like storage, memory and computing capabilities in running VMs.

Following the use case of Section~\ref{sec:usecase}, the next policy deals with the \textbf{first concern} by increasing RAM in the VM with CPU-supported ASD module \textit{(cpuASD)} when the usage of memory is exceeding a given threshold \textit{(RamUsageMax)}.

\noindent\small
\begin{flushleft}
	\begin{framed}
		\hspace{-0.5mm}Type(\#VI) $\wedge$ CPUSupportedASD(?cpuASD) $\wedge$ hasVM(?cpuASD,?VM) $\wedge$ integer[ramUsage~in~\#RamUsageMax]~hasRamUsage(?VM) $\rightarrow$ increaseRam(?VM)
	\end{framed}
\end{flushleft}
\normalsize

\subsection{Anomaly Detection Function Policy}

Anomaly Detection Function (ADF) policies are aimed at deploying and configuring automatically the different AD models proposed by our solution. Among the possible configurations, we highlight changing the model input, updating the detection model, or modifying the model parameters.

Taking into account the proposed use case, the next anomaly detection policy manages the \textbf{second concern} by updating the ASD model running in RAN. To attain this, the next policy detects when a new and improved model \textit{(newModel)} is released and updates the GPU-supported ASD module \textit{(gpuASD)} running in RAN with the new model.

\noindent\small
\begin{flushleft}
	\begin{framed}
		\hspace{-0.5mm}Type(\#ADF) $\wedge$ ASDModel(?newModel) $\wedge$ GPUSupportedASD(?gpuASD) $\wedge$ hasModel(?gpuASD, ?currentModel) $\wedge$ isImprovedVersion(?newModel, ?currentModel) $\rightarrow$ updateModel(?gpuASD,?newModel)
	\end{framed}
\end{flushleft}
\normalsize

\subsection{Mobile Edge Application Policy}
\label{sub:MEApp}

Mobile Edge Application (MEApp) policies are able to manage the ME applications in real-time and on-demand, particularly, those geared towards detecting and reacting to network anomalies. By using these policies, our proposal is capable of managing and configuring the internal behaviors of ME applications dynamically. Furthermore, these policies allow instantiating/dismantling applications as well as migrating or balancing their load work to optimize their use.

Following our use case, the next policy deals with the \textbf{third concern} by deploying a new application with a GPU-supported ASD module \textit{(gpuASD)} when the number of flows \textit{(numNetFlows)} managed by the CPU-supported ASD \textit{(cpuASD)} is exceeding a given threshold \textit{(NetFlowsMaxForCPU)}.

\noindent\small
\begin{flushleft}
	\begin{framed}
		\hspace{-0.5mm}Type(\#MEApp) $\wedge$ CPUSupportedASD(?cpuASD) $\wedge$ isLocated(?cpuASD,?RAN) $\wedge$ integer[numNetFlows~in~\#NetFlowsMaxForCpu]~hasNumNetFlows(?cpuASD) $\rightarrow$ GPUSupportedASD(?gpuASD) $\wedge$ deployMEApp(?gpuASD,?RAN)
	\end{framed}
\end{flushleft}
\normalsize

Finally, the next policy resolves the \textbf{fourth concern} by deploying and configuring a DPI tool like Snort (snortC\&C) when an anomaly based on a suspicious C\&C channel is detected. It should be borne in mind that a botnet could involve several RANs and, therefore, a DPI tool will be deployed in each one.

\noindent\small
\begin{flushleft}
	\begin{framed}
		\hspace{-0.5mm}Type(\#MEApp) $\wedge$ hasOutput(\#NAD, ?output) $\wedge$ isSuspiciousC\&C(?output) $\wedge$ hasLocation(?output,?ran) $\rightarrow$ Snort(?snortC\&C) $\wedge$ deployMEApp(?snortC\&C,?ran)
	\end{framed}
\end{flushleft}
\normalsize

Other stricter policy actions could be taken by this type of policy, such as drop or reject suspicious network flows through the configuration of traffic control service of the ME platform in the RAN.


\section{Use case orchestration}
\label{sec:orchestration}

We present the orchestration process for our use case showing the internal steps involved in making policy actions into the proposed architecture.

\begin{figure}[hbtp]
	\centerline{\includegraphics[width=0.8\columnwidth]{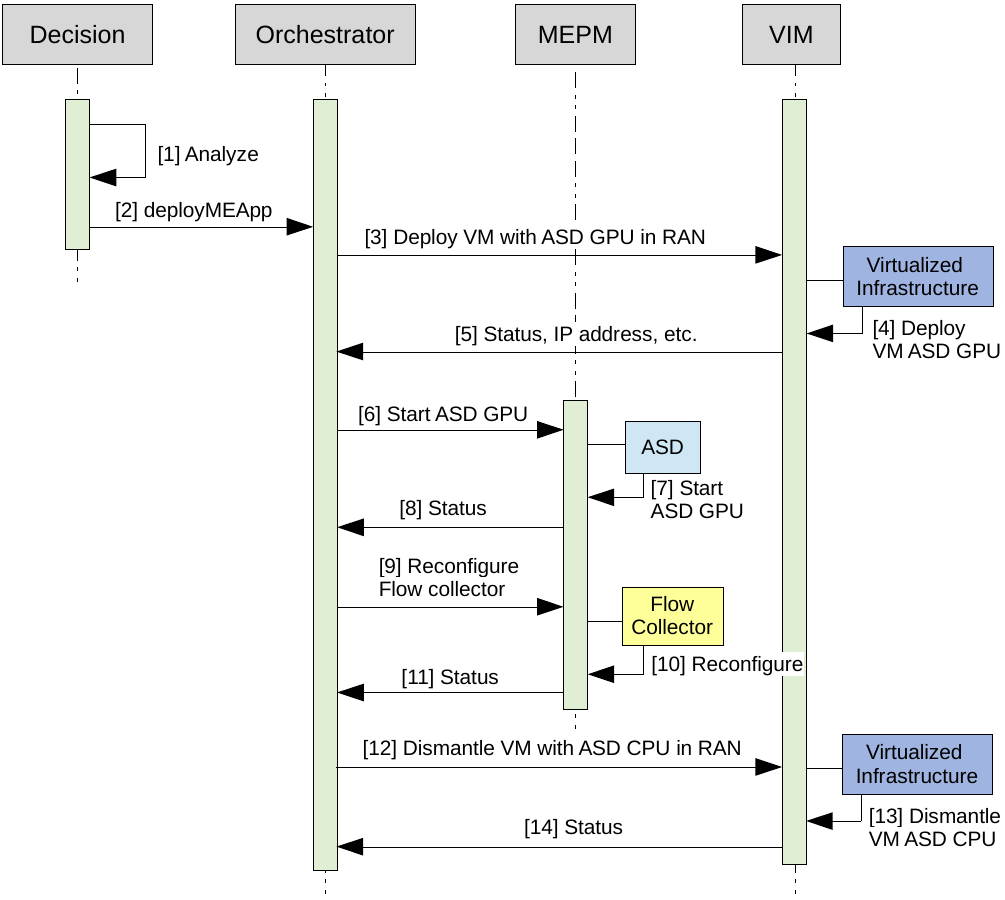}}
	\caption{Sequence diagram of the MEApp policy enforcement.}
	\label{fig:diagram}
\end{figure}

Following the use case defined in Section~\ref{sec:usecase}, we introduce below the internal steps performed by the proposed architecture. The aim is to enforce the reaction considered by the first MEApp policy detailed in Section~\ref{sub:MEApp}. In this context, \figurename~\ref{fig:diagram} shows the different steps performed by the elements of the architecture to deploy and configure a new application with a GPU-supported ASD application in the RAN.

The Decision element evaluates continuously the state of the network as well as the management policies predefined by the administrator (step 1 of \figurename~\ref{fig:diagram}). In this context, once the conditions defined by the MEApp policy are accomplished, the Decision element decides to deploy a new VM with the GPU-supported ASD in the RAN. This decision is sent to the Orchestrator (step 2), which interacts with the Virtualisation Infrastructure Manager (VIM) to instantiate, in the RAN, the image of the VM with the GPU-supported ASD application already installed (step 3). The VIM checks that the available physical resources (processing, memory, and storage) are sufficient to perform the instantiation of the VM and instantiates it in the Virtualisation Infrastructure (step 4). After that, it notifies the Orchestrator about the current state of the VM by providing specific information of the VM such as its IP address (step 5). Once the Orchestrator knows that the VM has been deployed and is running, it sends the MEP Manager the VM information as well as the order to start the GPU-supported ASD application (step 6). The MEP manager starts the GPU-supported ASD application thanks to the information of the deployed VM provided by the Orchestrator (step 7). After receiving the OK status of the GPU-supported ASD application (step 8), the Orchestrator communicates again with the MEP Manager in order to re-configure the Flow Collector service (step 9). This re-configuration consists in sending the output of the Flow Collector application to the new GPU-ASD application instead of the previous one (step 10). Finally, once the Orchestrator knows that the reconfiguration has been made (step 11), it communicates with the VIM to dismantle the existing VM with the previous CPU-supported ASD (step 12). After that, the VIM dismantles the VM (step 13) and it responds with a positive message to the Orchestrator (step 14).


\section{Deployment and performance results}
\label{sec:performance}

\begin{figure*}[!ht]
	\centerline{\includegraphics[width=1.0\textwidth]{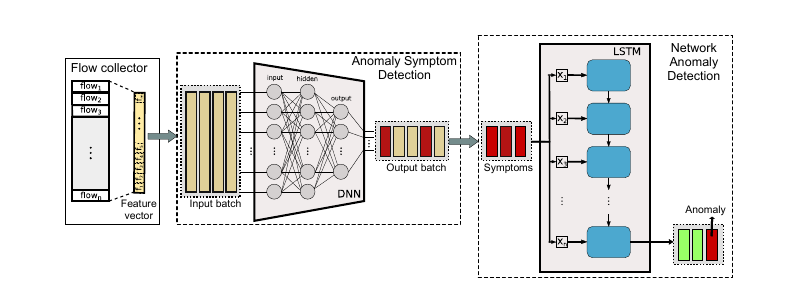}}
	\caption{Detail of the implementation of ASD and NAD modules.}
	\label{fig:ourmodule}
\end{figure*}

For the implementation of the detection modules, we propose to use deep learning techniques in order to analyze network flows. \figurename~\ref{fig:ourmodule} depicts the main components of the proposed AD system.

In our AD system, the flow collector gathers all the different flows during a given period of time and calculates a vector of features that the ASD module will classify as either anomalous or normal. If an anomaly is suspected, a symptom packet made up of the feature vector involved, a timestamp, and the type of anomaly detected (when a multi-class method is being used) are sent to the next level, the NAD module. The NAD receives several streams of symptoms from the ASDs, it sorts them by their timestamps, and assembles a time sequence of symptoms. The task of deciding whether this sequence belongs to any of the attack categories can be seen as a sequence-of-symptoms classification problem.

Due to the high volume of traffic that each RAN has to manage in a 5G network, it is crucial that the ASD classification has to be made as quickly as possible, even sacrificing accuracy for a lower response time. However, this must be made consuming as few resources as possible. There exists a variety of available deep learning frameworks that offer flexible development environments. Likewise, they provide a set of highly optimized libraries that allow building and deploying deep learning models efficiently. The runtime performance of a subset of those frameworks is studied and presented in \mycitep{loren2018access}. Our runtime study focuses on the ASD module because it is the most resource-demanding part of our system. Note that ASD module has to analyze every feature vector in networks with extremely high bandwidth.

It is important to mention that we do not require a high level of accuracy during the ASD classification because we rely on the second level (NAD) to refine the final detection results. Considering these requirements, Deep Neural Networks (DNN) and Long Short-Term Memory (LSTM) Recurrent Networks have been the models chosen for ASD and NAD respectively.

\subsection{ASD classification performance}
\label{sub:ASDclassperf}

From the classification accuracy point of view, it is necessary to figure out whether good detection/classification results can be obtained with this proposal. We used a DBN followed by a classification layer as the DNN. The tests performed have demonstrated that this model is capable of yielding successful experimental results.

The 5G scenario is rather complex, and we need a publicly available dataset suitable to be used in order to test our proposal. In this sense, CTU \mycitep{garcia2014empirical} is an appropriate dataset to be used in this context. It tries to accomplish all the requirements to be considered as a effective dataset: it has real botnets attacks and not simulations, unknown traffic from a large network, ground-truth labels for training and evaluating, as well as different types of botnets. The CTU dataset comprises thirteen scenarios with different numbers of infected computers and seven botnet families \mycitep{garcia2014empirical}.

An efficient machine learning mechanism also needs a set of highly discriminative features. Deep learning has proved its ability to extract high-level features from raw data automatically. To take advantage of this fact, our feature vector is made up of a variety of statistical measures and information metrics computed from a batch of network flows, as described in \mycitep{loren2018access}. This batch is created after receiving a certain number of flows (offset). By using deep neural networks, it is expected that the model will learn increasingly higher-level features by combining the original input components, thus reducing the necessity of feature engineering.

In our study, two different training/test partitions were made. On the one hand, our first dataset was built by splitting the original CTU dataset into training (80\%), and test (20\%), both containing samples of every botnet. Therefore, the trained models were evaluated with the same botnets they already knew. On the other hand, our second dataset was built using the same partition of the thirteen scenarios that make up the CTU dataset. The scenarios were split into training and test, meeting the following constraints: the training datasets should be approximately 80\% of the whole dataset, the testing dataset should be approximately 20\%, and none of the botnet families used in the training dataset should be used in the testing dataset (i.e. training with scenarios 3, 4, 5, 7, 10, 11, 12, 13 and testing with 1, 2, 6, 8 and 9) \mycitep{garcia2014empirical}. This ensures generalization and detection of new behaviors, a critical aspect when dealing with cyber defense scenarios in 5G networks.

We restricted our search domain to three DBNs with up to 6 hidden layers and ReLu as activation function, an input vector of 288 features, batch normalization and a binary classifier as output layer (anomaly vs. normal). The activation function of the output was a sigmoid and cross-entropy was used as cost function. The hyperparameters of each of the given networks were: learning rate $\in [0.001, 0.5]$, dropout $\in [0, 0.4]$ and L2-regularization $\in[0,0.2]$. Hyperparameter tuning was carried out by means of 10-fold cross-validation and randomized search.

The chosen performance metrics have been precision and recall:

\[
Precision = \frac{True~positives}{True~positives+False~positives}
\]

\[
Recall = \frac{True~positives}{True~positives+False~negatives}
\]

These metrics are more informative than accuracy when the datasets are highly unbalanced. This is typically the case of anomaly related datasets.

When trained and tested with all the botnets, the 128-64-32 model achieved the highest F1-score, with a precision of $0.9537$ and a recall of $0.9954$; that is, $95.37\%$ of the anomaly predictions were actually anomalies, and $99.54\%$ of the actual anomalies were correctly classified (see \tablename~\ref{tab:knownNN}). Regarding the models tested with unknown scenarios, the 16-8-4 model obtained the better generalization results in spite of its simplicity. The evaluation of the trained model reached a precision of 0.6861 and a recall of 0.7096 on average. Approximately 71\% of the anomalous traffic was correctly detected, even though the model had never seen those botnets before. The global scores of each model evaluated on the unknown scenarios are presented in \tablename~\ref{tab:unknownNN}. In addition, a breakdown by unknown scenario of the scores obtained by the 16-8-4 model is provided in \tablename~\ref{tab:scenarios}.

\begin{table}[htbp]
	\centering
	\caption{Classification results of the three selected deep neural networks regarding known botnets.}
	\begin{tabular*}{0.76\columnwidth}{lccccc}\hline\\[-1em]
		Hidden layers & LR & $\lambda_{L2}$ & \footnotesize{Precision} & Recall & F1-score\\\hline\\[-1em]
		16-8-4 & 0.01 & 0.00 & 0.8311 & 0.9947 & 0.9055\\
		128-64-32 & 0.01 & 0.00& \textbf{0.9537} & \textbf{0.9954} & \textbf{0.9741}\\
		\footnotesize{128-128-64-64-32-32} & 0.1 & 0.01 & 0.9534 & 0.9904 & 0.9715\\\hline\hline
	\end{tabular*}
	\label{tab:knownNN}
\end{table}

\begin{table}[htbp]
	\centering
	\caption{Classification results of the three selected deep neural networks regarding unknown botnets.}
	\begin{tabular*}{0.76\columnwidth}{lccccc}
\\\hline\\[-1em]
		Hidden layers & LR & $\lambda_{L2}$ & \footnotesize{Precision} & Recall & F1-score\\\hline\\[-1em]
		16-8-4 & 0.01 & 0.00 & 0.6861 & \textbf{0.7095} & \textbf{0.6976}\\
		128-64-32 & 0.01 & 0.01& 0.6910 & 0.6775 & 0.6842\\
		\footnotesize{128-128-64-64-32-32} & 0.01 & 0.01 & \textbf{0.8693} & 0.4803 & 0.6187\\\hline\hline
	\end{tabular*}
	\label{tab:unknownNN}
\end{table}

\begin{table}[htbp]
	\centering
	\caption{Breakdown of classification results by unknown scenario for the 16-8-4 model.}
	\begin{tabular*}{0.50\columnwidth}{lccc}\hline\\[-1em]
		Test set & Precision & Recall & F1 score\\\hline\\[-1em]
		Scenario 1 & 0.742 & 0.902 & 0.814\\
		Scenario 2 & 0.631 & 0.544 & 0.584\\
		Scenario 6 & 0.724 & 0.937 & 0.812\\
		Scenario 8 & 0.410 & 0.756 & 0.531\\
		Scenario 9 & 0.941 & 0.388 & 0.549\\\hline\hline
	\end{tabular*}
	\label{tab:scenarios}
\end{table}

Our expectations are that these results will improve reasonably by the second level detector in the NAD module. This module will build a timestamped anomaly sequence from the symptom information received from the different RANs. This sequence will then be analyzed for botnet complex pattern detection and filtering of the misclassified anomalies.

\subsection{ASD model evaluation throughput}
\label{sub:ASDthroughput}

From the prediction time point of view, ASD requires a deep-learning framework that evaluates the trained model as quickly as possible due to the time restrictions inherent in 5G networks. There is a significant number of libraries and frameworks related to deep learning that can be used to train the model proposed in this article. For our experiments we have selected the following frameworks: TensorFlow r1.4 (Abadi et al \citeyear{abadi2016tensorflow}) and Caffe2 r0.8.1 \mycitep{caffe2}. Both frameworks achieve a high performance on CPU and GPU respectively and they have been widely used in different R\&D as well as commercial projects.

In this experimental work, our aim is to test a sufficient complex model which can stress the ASD module. In this sense, the throughput of the previous frameworks has been evaluated by using a DBN architecture made up of six hidden full layers followed by a output binary classification layer (128-128-64-64-32-32-1). An input feature vector of 256 elements has been selected instead of 288 features for performance purposes. All the test were carried out using a workstation with 32GB RAM, a six-core Intel i7-5930K at 3.5GHz with hyper-threading running Ubuntu Linux 16.04, and one NVIDIA GeForce GTX 1080 with 8GB RAM.

\begin{figure*}[!htb]
	\centering
	\subfloat[Evaluation throughput]{\includegraphics[width=0.48\textwidth,height=4cm]{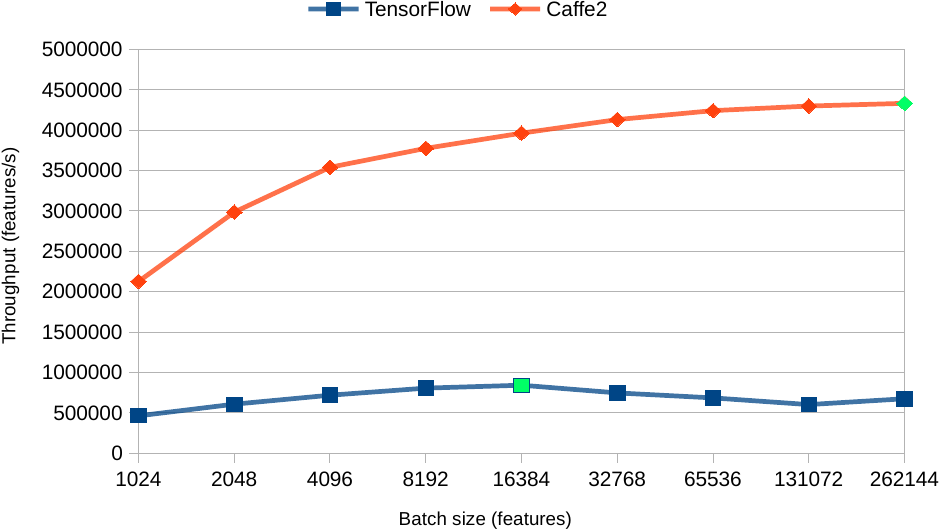}\label{subfig:throughput}}\quad
	\subfloat[Batch evaluation time ($t_{ev}$)]{\includegraphics[width=0.48\textwidth,height=4cm]{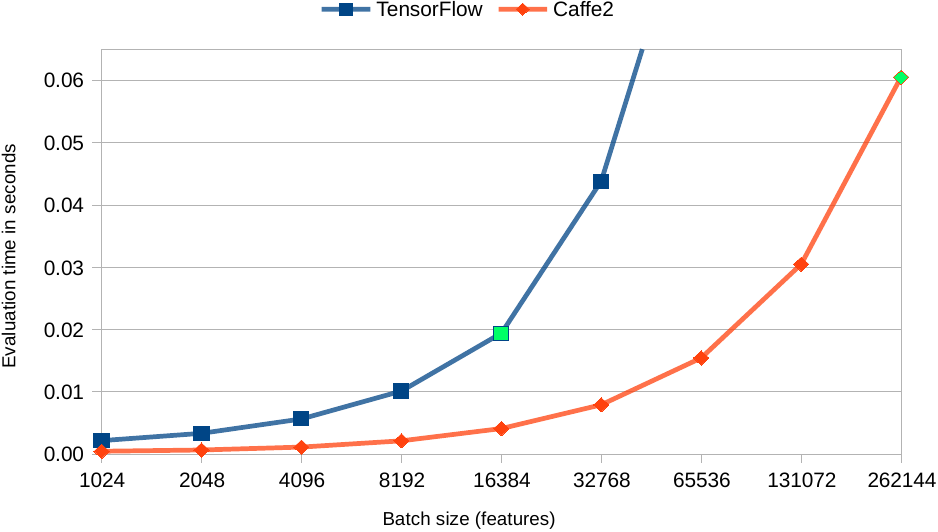}\label{subfig:batchevaluation}}
	\caption{Evaluation throughput and batch evaluation runtime ($t_{ev}$) of TensorFlow on CPU and Caffe2 on GPU for a wide range of feature batch sizes. The green markers point out the optimum values.}
	\label{fig:throughput}
\end{figure*}

In our tests with a set of deep learning frameworks, TensorFlow had the best performance evaluating our model on CPU. Conversely, Caffe2 showed the best results when using GPU. In \figurename~\ref{fig:throughput}(a), evaluation rate of feature vectors w.r.t. batch size is plotted. TensorFlow reaches its maximum throughput with a feature batch size of 16\,384 whereas Caffe2 improves its performance as the batch size increases. Therefore, if Caffe2 is used on a GPU, the batch size as large as possible should be chosen. The runtime of each framework when evaluating a batch of a given size is shown in \figurename~\ref{fig:throughput}(b).

\subsection{Anomaly detection time estimation}

A relevant point to evaluate our proposal is an estimation of the response time of our system when it is stressed with a high flow rate. In other words, we are interested in studying the detection time lag as a function of the number of flows per second. Initially, we consider that an upper limit of time has to be set so as to force an evaluation when the network traffic rate is rather low. The time needed to fill a batch would be excessive even if a relatively small batch size is used. For illustration purposes, we have chosen a limit ($t_{limit}$) of 5 seconds.


Equation~\eqref{eq:featuretime} defines the mean time elapsed between features ($t_{bf}$) given the number of flows per second collected by a RAN, and the value of the offset. Likewise, \eqref{eq:filltime} gives an expression of how long it takes to fill a feature batch ($t_{fill}$).

\begin{equation} \label{eq:featuretime}
t_{bf} = \frac{\textit{offset}}{\textit{flow rate}}
\end{equation}

\begin{equation}\label{eq:filltime}
t_{fill} = min(t_{limit},\textit{batch size}\times t_{bf}) = min(t_{limit},\frac{\textit{batch\ size}}{\textit{flow\ rate}} \times \textit{offset})
\end{equation}

When added to the model evaluation time for the batch ($t_{ev}$), shown in \figurename~\ref{fig:throughput}(b), we obtain the lower and upper bounds of the elapsed time between the reception of an anomaly and the detection: $t_{det} \in [t_{bf}+t_{ev}, t_{fill}+t_{ev}]$

\begin{figure*}[!ht]
	\centering
	\includegraphics[width=1.0\textwidth]{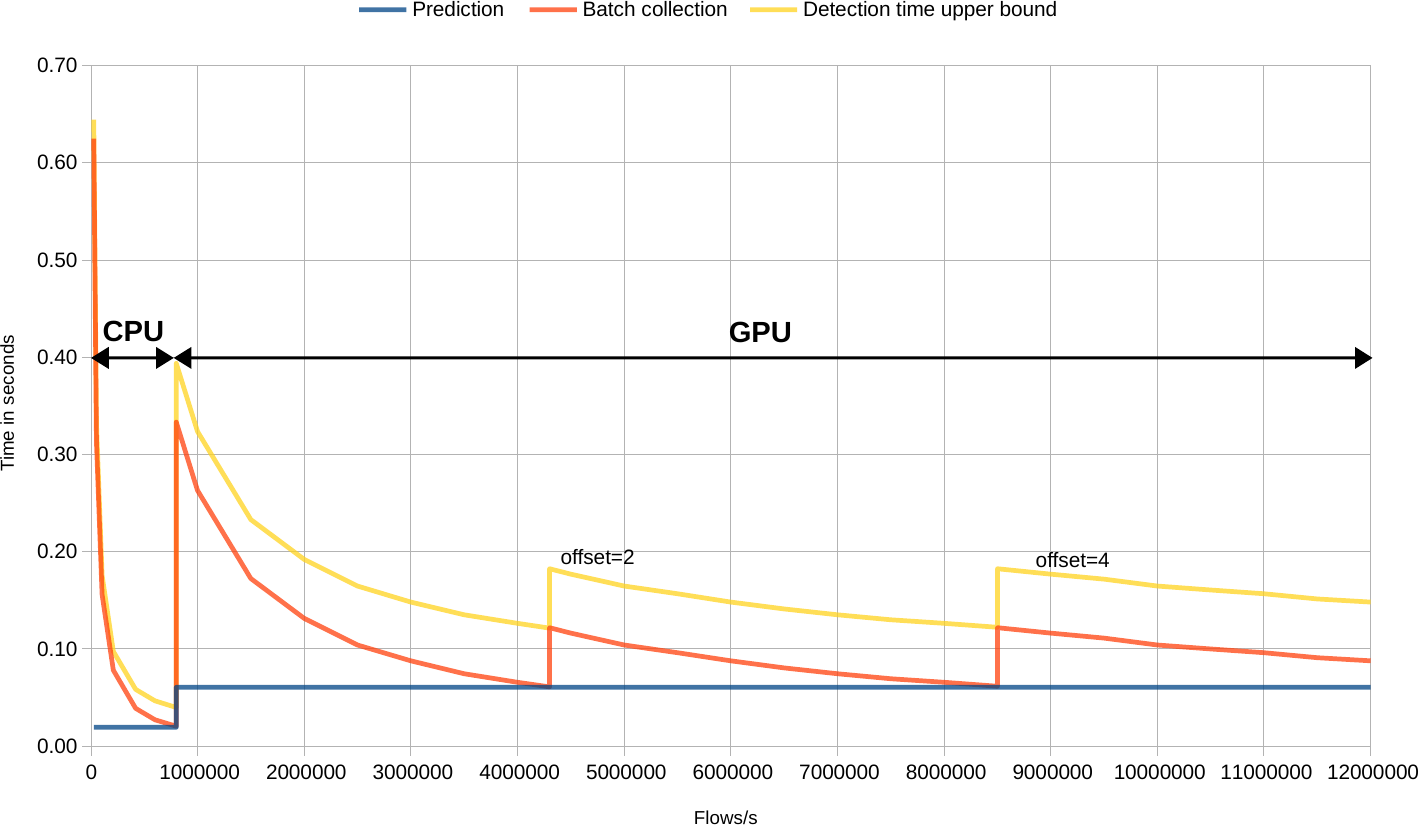}
	\caption{Detection time upper bound when the rate of flows per second is increased. Red line: time needed to fill a batch ($t_{fill}$); blue line: runtime spent on evaluating the batch ($t_{ev}$); yellow line: total time lag between the reception of the first feature in the batch and the computation of the batch prediction ($t_{fill}+t_{ev}$). TensorFlow is used in the CPU VM and Caffe2 in the GPU VM.}
	\label{fig:detectiontime}
\end{figure*}

An anomaly detection time upper bound is plotted in \figurename~\ref{fig:detectiontime}. As long as the flow rate is low, a CPU is sufficient to evaluate the prediction model for each feature. Assuming TensorFlow as the framework used in this case, our best choice would be a batch size of 16\,384 features. If our 5-second collection time limit is reached, a batch evaluation is forced. This case is not shown in \figurename~\ref{fig:detectiontime} because it is an uninteresting condition. However, when the flow rate increases, the 16\,384-batch will be quickly filled and an anomaly prediction will be computed, with a runtime of 0.0194 seconds. This can be observed in the blue line of the same figure. The red line represents the batch filling time according to~\eqref{eq:filltime}, while the yellow one does the sum of the two values. TensorFlow over CPU can be used as long as $t_{fill} \geq t_{ev}$.

The equality is reached at 842\,600 flows per second. At this point we have two options: we could either set an offset of 2 flows, halving the feature generation rate or request a new virtual machine with GPU support and the framework Caffe2 running on it. This second option is depicted in \figurename~\ref{fig:detectiontime}.

When the new VM is deployed, a new batch size of 262\,144 features is selected to take advantage of the GPU's higher throughput. This corresponds in the figure to a step in the blue line to a new evaluation time of 0.060 seconds, and a sudden increase of the batch filling time (red line), starting a gradual drop as the flow rate grows.

Again, this configuration is suitable as long as the time of filling the batch is greater than the evaluation time, and both curves meet at a rate of 4\,332\,000 flows per second. In this situation we can set an offset of 2 flows and, as a result, the flow rate is halved and the red line declines more gently. A light increase of the offset results in a new enlargement of the operating range. This can be repeated by enlarging the offset as necessary. In the plot the offset is doubled, allowing a flow rate of up to approximately 17\,000\,000 flows per second.

Given the reasonable assumption that all positions in the batch are equally likely, the average detection time is the mean of a uniform distribution between $t_{bf}+t_{ev}$ and $t_{fill}+t_{ev}$, that is, $({t_{bf}+t_{fill}})/2 + t_{ev}$. In our 5G context, $t_{bf}$ can be considered negligible in virtually all cases. Therefore the average detection time in~\figurename~\ref{fig:detectiontime} would be the middle point between the blue ($t_{ev}$) and red ($t_{fill}+t_{ev}$) curves.

\begin{figure*}[!ht]
	\centerline{\includegraphics[width=0.9\textwidth]{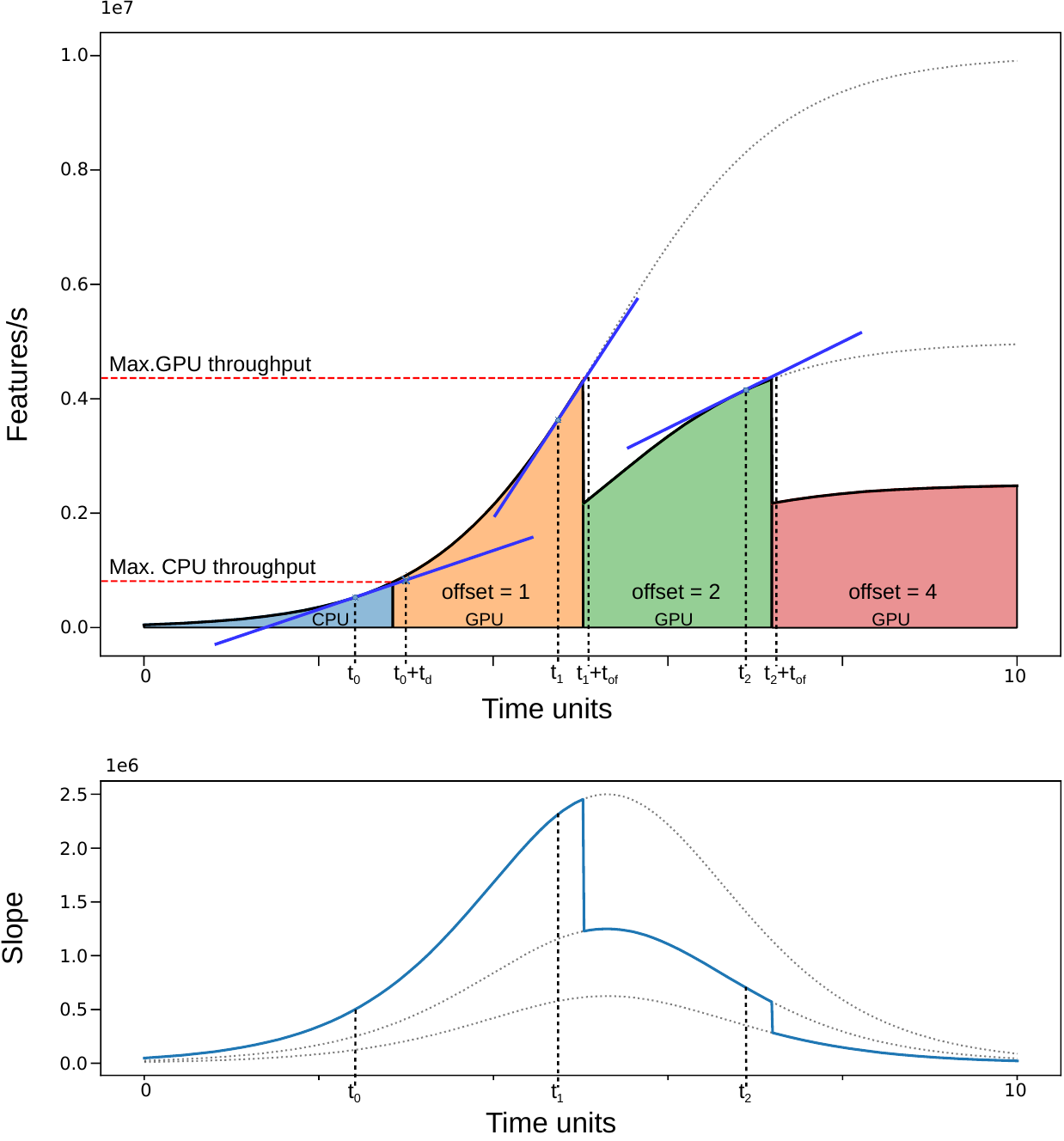}}
	\caption{Evolution of the framework deployed and the collector offset selected during a large movement of people from one RAN to another. Each offset has its corresponding dotted function shape. \emph{Above:} evolution of the deployed CPU/GPU-based VM and collector offset over time. Features per second are supposed to follow~\eqref{eq:sigmoidfeatures}, $t_d$ is the time needed to deploy a new VM and $t_{\textit{of}}$ is the time needed to change the offset. The tangents show how the first derivative estimation can anticipate the tendency. The offset is doubled in each increment. \emph{Below:} first derivative of each sigmoid. The slope at $t_0$ can be used to obtain a prediction of the curve value at $t_0+t_d$ with the linear approximation $f(t_0+t_d) \approx f(t_0)+f'(t_0)\times t_d$.}
	\label{fig:sigmoids}
\end{figure*}

\subsection{ASD system dynamics analysis.}

In order to set up a suitable context in which to show the dynamic behavior of our ASD, we establish the following premises:

\begin{enumerate}
	\item Initially, there is an extremely low network traffic rate (i.e. 50\,000 flows/s) in a RAN.
	\item Events make large amounts of users come from other places, becoming connected to our RAN.
	\item The arrival rate of a large amount of users to an event can be modeled as a Gaussian function, and the integral of this family of functions has a sigmoid shape.
	\item If an average number of 10 flows/s per user is assumed, and the number of user devices goes from 5\,000 to 1\,000\,000 in 10 time units, then the number of flows per second w.r.t. time (in the same time units) can be modeled as a sigmoid function that starts at 50\,000 flows/s, saturates at 10\,000\,000 flows/s, and reaches the 99\% of this value at $t=10$, namely,
		\begin{equation} \label{eq:sigmoidflows}
			flows(t) = \frac{10^7}{1+e^{-(t-5.2933)}}
		\end{equation}

		\begin{equation} \label{eq:sigmoidfeatures}
			features(t) = \frac{flows(t)}{\textit{offset}}
		\end{equation}
\end{enumerate}

The shape of the sigmoid function and its derivative for three different offsets is plotted in \figurename~\ref{fig:sigmoids}. The graph presents the dynamic behavior of our system over time in this context. Initially our system has a small feature rate that can be easily managed by a VM with a deep learning framework based on CPU (in our case, TensorFlow). The RAN is continuously estimating the derivative of the feature rate function. Therefore, if a VM deployment time of $t_d$ time units is assumed, we can use the derivative to predict whether a new VM will be necessary in $t_d$ time units. If so, we can request the new VM ahead of time, so that it is active when the critical load is reached. Similarly, if we know the time that an offset increase needs to become effective, for example $t_{\textit{of}}$, we can request the change in advance. This is depicted in \figurename~\ref{fig:sigmoids}, where the blue tangent lines represent the traffic load prediction. The leftmost one predicts that our VM with TensorFlow will become overloaded in $t_d$ time units. Thus, our RAN requests a GPU-oriented VM that will be ready before the critical point is reached. The other tangents are used to request the offset increase at least $t_{\textit{of}}$ time units before it is needed.

\hfill

As a conclusion of this experimental results, the dynamic response of our system in a realistic scenario has been studied, and a seamless method for the enforcement of policies considering the VM deployment time has been proposed. In addition, the detection time of our ASD module has been estimated from the throughput of two well-known deep learning frameworks.


\section{Conclusions and future work}
\label{sec:conclusions}

In this article, we have presented a MEC-oriented architecture to manage the network anomaly detection by using policies. By means of three types of policies and an orchestration process in charge of making policy actions, our proposed architecture can deploy diverse actions to assure an effective anomaly detection process in real time. The functional capabilities of our proposal have been highlighted by a realistic use case where several concerns of 5G networks are shown.

Moreover, we have presented the deployment of the proposed detection system using deep learning techniques to analyze network flows and to detect network anomalies. Several experiments have been realized to evaluate the ASD classification performance, the ASD model evaluation throughput and anomaly detection performance. The experimental results have shown that our proposed architecture can provide effective anomaly detection and adapt the detection modules in real-time and in autonomic way.

As future work, we plan to train ASD and NAD modules with data obtained from a real 5G scenario and to evaluate the accuracy of the anomaly detection architecture as a whole. This has been identified as the next step in a real 5G network, as the one being currently built in the SELFNET EU H2020 project.


\section*{Acknowledgements}

This work has been partially supported by a S\'eneca Foundation grant within the Human Resources Researching Postdoctoral Program 2018, a postdoctoral INCIBE grant within the ``Ayudas para la Excelencia de los Equipos de Investigaci\'on Avanzada en Ciberseguridad'' Program, with code INCIBEI-2015-27352, the European Commission Horizon 2020 Programme under grant agreement number H2020-ICT-2014-2/671672 - SELFNET (\textit{Framework for Self-Organized Network Management in Virtualized and Software Defined Networks}), and the European Commission (FEDER/ERDF).


\bibliographystyle{spbasic}
\bibliography{references}

\end{document}